\documentclass[showpacs,amssymb,amsmath]{revtex4}
\usepackage{graphicx}
\usepackage{psfrag}
\usepackage{subfigure}

\newcommand{\ket}[1]{|#1\rangle}

\begin{document}

\title{Simulation of the Single- and Double-Slit Experiments with Quantum Walkers}
\author{A.C. Oliveira, R. Portugal}%
\email{portugal@lncc.br}%
\affiliation{Laborat\'{o}rio Nacional de Computa\c{c}\~{a}o
Cient\'{\i}fica, LNCC, C.P. 95113, Petr\'{o}polis, RJ,
25651-075, Brazil}%
\author{R. Donangelo}%
\affiliation{Universidade Federal do Rio de Janeiro, UFRJ, C.P.
68528, Rio de Janeiro, RJ, 21941-972, Brazil}%
\date{\today}%

\begin{abstract}
We employ the broken-link model to create a barrier with slits in a two-dimensional lattice.
The diffraction and interference patterns of the probability distribution of quantum walkers
passing through the slits are analyzed. Simulations were performed using the main types of
coins, and display diffraction and interference patterns that depend on the choice of coins.
\end{abstract}

\maketitle %

\section{Introduction}

The main models of quantum walks, known as discrete-time and continuous-time, were introduced
by Aharanov, Davidovich, and Zagury~\cite{Aharonov} and Farhi and Gutmann~\cite{Farhi},
respectively. Both of those models present radically new features when compared to the classical
random walk, one of the most remarkable is that quantum walks spread in a ballistic way, in
contrast with the diffusive behavior of classical random walkers.
This property motivated many studies pursuing to introduce quantum algorithms much faster than
their classical counterparts~\cite{Farhi,Shenvi,Ambainis}.

Many papers in the recent literature address the theory of one-dimensional quantum walks in order
to understand both its physical properties and possible applications~\cite{Aharonov,Nayak}.
However, in the general case, there are some properties that appear only in two or higher dimensions.
In our previous work on decoherence in two-dimensional quantum walks~\cite{Oliveira}, we have shown
that, when increasing the decoherence rate in an appropriate way, there is a transition from a
decoherent two-dimensional to a coherent one-dimensional quantum walk. To obtain this result,
we have employed the broken-link decoherence method introduced by Romanelli \textit{et al.} for the
one-dimensional case~\cite{Romanelli}.
In order to extend the broken-link method to higher dimensional walks, in~\cite{Oliveira} we have
generalized the evolution equation to generic broken-link topologies and lattices of any dimension.

As shown in the present work, that generalized evolution equation can also be used to study quantum
walks with rather arbitrary boundary conditions. The freedom to choose the boundary conditions allows
to study random walks on a large variety of topologies. In order to illustrate the method, we study
here the behavior of quantum walkers passing through one and two slits. The standard double-slit
experiment using light waves (Young's experiment) proved conclusively the wave nature of light~\cite{Born}
and has notedly been employed by Feynman to illustrate the differences between the classical and quantum
worlds~\cite{Feynman}. We found interesting to ascertain, in the case of quantum walkers, which as noted
possess ballistic properties, but are not characterized by a single wavelength, whether they would display
diffraction and interference patterns, or would behave more as classical particles.

\begin{figure}[h]
    \includegraphics[width=6.2cm,scale=1]{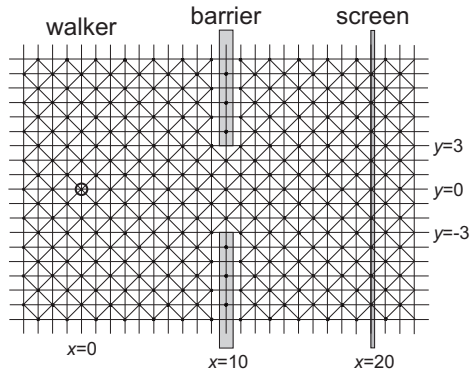}%
    \caption{General setup of the simulation at $t=0$.
    The walker is in the node $x=y=0$. At $x=10$ there is a barrier with
    one slit of 3 units wide. The barrier is made of a line of
    sites with the links broken. At $x=20$ there is a screen.}
    \label{DoubleSlitFig}
\end{figure}

Our analysis is based on the simulation of different kinds of
quantum walkers moving on a region of the two-dimensional lattice
illustrated on Fig.~\ref{DoubleSlitFig}. Notice that, as in
Ref.~\cite{Oliveira}, the lattice links are parallel to the main
or secondary diagonals of the grid. The walker starts at one point
on the lattice, indicated by an open circle, and moves due to the
successive application of the (unitary) evolution operator, as is
usual in the theory of quantum walk. At some point not very close
to the origin of the evolution, there is a barrier perpendicular
to the horizontal axis with either one or two slits. The barrier
consists of a succession of sites with broken links between them,
while the slits correspond to sites with working links. At some
distance from the barrier there is a screen, also perpendicular to
the horizontal axis, which accumulates the value of the
probability as the walker passes. The general setup is similar to
the standard Young's double-slit interference experiment.

Nowadays, double-slit experiments are performed such that only one particle passes at a time, proving that
quantum mechanics describes correctly the one-particle interference pattern. The same strategy is used in
the present simulations. We analyze the interference pattern in the screen by considering the equation for
the evolution of a single quantum walker. From the analysis of the probability distribution we predict what
would happen in actual experiments.

We have organized the paper as follows. In Sec.~\ref{sec:2dlattice} we review the
equations that govern the evolution of a quantum walker in 2-D lattices with broken
links. The diffraction and interference patterns of walkers passing through one slit
is analyzed Sec.~\ref{sec:1slit} and the double-slit case in Sec.~\ref{sec:2slits}.
A general discussion with conclusions and suggestions for further studies is presented
in the last section of this work.
\section{Quantum Walks with Boundary Conditions}
\label{sec:2dlattice}

We review here the main elements needed to study the evolution of a quantum walker moving
on a 2-D lattice with broken links. Full details of this treatment may be found in
Ref.~\cite{Oliveira}.

A coined quantum walk on an infinite two-dimensional lattice takes place on a Hilbert
space $\mathcal{H}_4\otimes \mathcal{H}_\infty$, where $\mathcal{H}_4$ is the coin
space and $\mathcal{H}_\infty$ is the lattice space. The coin consists of two qubits
with basis $\{{\left |j,k\right \rangle}, j,k\in\{0,1\}\}$. As mentioned above, we
consider that the links are along either the main or the secondary diagonals of the
lattice. Thus, the basis for $\mathcal{H}_\infty$ is
$\{{\left |m,n\right \rangle}, m,n\,{\rm integers}\}$ such that $m+n$ is even.

The generic state of the quantum walker is given by the wavefunction
\begin{equation}
{\left |\psi(t)\right \rangle}  =\sum_{j,k=0}^{1}
\sum_{m,n=-\infty}^{\infty}A_{j,k;\,m,n}(t){\left |{j,k}\right
\rangle}{\left |m,n\right \rangle}. \label{eq:estgeral2d}
\end{equation}
The evolution operator for a single step of the walk is
$U=S \circ(C\otimes I)$, where
\begin{equation}
C{\left |j',k'\right \rangle}=\sum_{j,k=0}^{1}
C_{j,k;\,j',k'}{\left |j,k\right \rangle} \label{coin2d}
\end{equation}
is the coin operator, $I$ is the identity matrix, and $S$ is the shift operator given by
\begin{eqnarray}
S{\left |j^\prime,k^\prime\right \rangle}{\left |m,n\right
\rangle} = \sum_{j,k=0}^{1}& &
\left[1-\delta_{j,j^\prime}-(-1)^{j^\prime}\right.\left.{\mathcal{L}_1}(1-j,1-k;
\,m+(-1)^j\delta_{j,j^\prime},n+(-1)^k\delta_{k,k^\prime} )\right]\times \nonumber\\%
& &\left[1-\delta_{k,k^\prime}-(-1)^{k^\prime}\right. \left.
{\mathcal{L}_2}(1-j,1-k;\,m+(-1)^j\delta_{j,j^\prime},n+(-1)^k\delta_{k,k^\prime})\right]\times \\
& &
\ket{j,k}\ket{m+(-1)^j\delta_{j,j^\prime},n+(-1)^k\delta_{k,k^\prime}},\nonumber
\end{eqnarray}
where $\mathcal{L}_1$ and $\mathcal{L}_2$ are two auxiliary functions required to specify
the broken links, one for each direction. These functions are described by
\begin{equation}
\mathcal{L}_1(j,k;\,m,n) =
\begin{cases}
(-1)^j, & \mbox{if  {\rm link} to  site $m+(-1)^j$,}
\mbox{$n+(-1)^k$ is closed,}\\
  0,    & \mbox{if  {\rm the link}  is open,}
\end{cases}\label{eq:link1}
\end{equation}
and
\begin{equation}
\mathcal{L}_2(j,k;\,m,n) =
\begin{cases}
(-1)^k, & \mbox{if  {\rm link} to  site $m+(-1)^j$,}
\mbox{$n+(-1)^k$ is closed,}\\
  0,    & \mbox{if  {\rm the link}  is open,}
\end{cases}\label{eq:link2}
\end{equation}
where $ j,k\in \{0,1\}$.

When the links are unbroken, the walker moves along the main diagonal, if the value of the
coin is ${\left |0,0\right \rangle}$ or ${\left |1,1\right \rangle}$, or along the secondary
diagonal if the value of the coin is ${\left |0,1\right \rangle}$ or ${\left |1,0\right \rangle}$.
If some links are broken, the walker moves, or stands still, in such a way that the probability
flux is conserved and the evolution is unitary. This is assured by the evolution
equations~\cite{Oliveira}:

\begin{eqnarray}
A_{1-j,1-k;\, m,n}(t+1)=\sum
_{{j^\prime,\,k^\prime}=0}^{1}C_{j+\mathcal{L}_1 \left(j,k;\,m,n
\right) ,k+ \mathcal{L}_2 \left(j,k;\,m,n \right)
;\,{j^\prime},{k^\prime}}\,A_{ {j^\prime},{k^\prime}; m+\mathcal{L}_1
\left(j, k;\,m,n \right) ,n+\mathcal{L}_2 \left(j,k;\,m,n
\right)}(t). \label{eq:evolution}
\end{eqnarray}

In the next sections, we analyze the diffraction and interference patterns produced by quantum
walkers passing through one and two slits. We use the Hadamard, Fourier, and Grover coins in the
simulations. The initial conditions are the same used in Ref.~\cite{Oliveira}, except for the
Hadamard walker, which is
 \begin{equation*}
{\left |\psi(0)\right \rangle} =\frac{1}{2}({\left |0\right
\rangle}+i{\left |1\right \rangle})({\left |0\right
\rangle}+i{\left |1\right \rangle}) {\left |0,0\right \rangle}.
\label{had0}
\end{equation*}
The analysis is based on the probability distribution, which is given by
\begin{equation}
P_{m,n}(t)=\sum_{j,k=0}^{1}{\left |A_{j,k;\,m,n}(t) \right |}^2.
\label{eq:e52d}
\end{equation}
As in the one-dimensional case, there is a parity pattern in the probability distribution in the
2-dimensional case, when there are no broken links. If the probability at one site of the lattice
is not zero, the probability at the neighboring sites should be zero.
When there are broken links, as in the slitted barrier, the reflections on the incidence side will
destroy this parity pattern. However, the wave that passes through the slit(s) conserves the parity
pattern. To clarify the description we shall plot only the non-zero probabilities.

\section{The single-slit case}\label{sec:1slit}

\begin{figure}[h]
    \includegraphics[width=5.5cm,angle=270,scale=1]{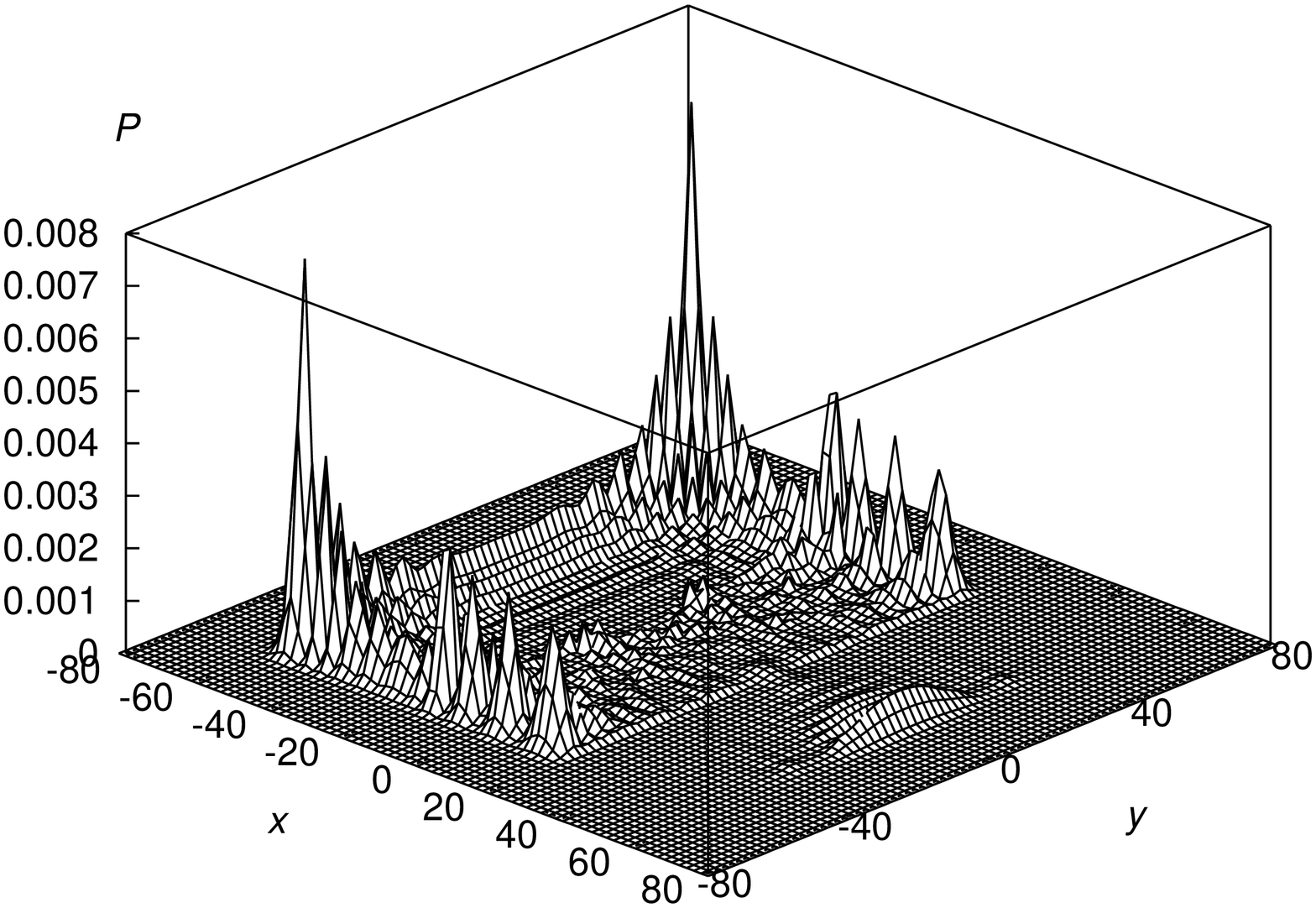} %
    \includegraphics[width=5.5cm,angle=270,scale=1]{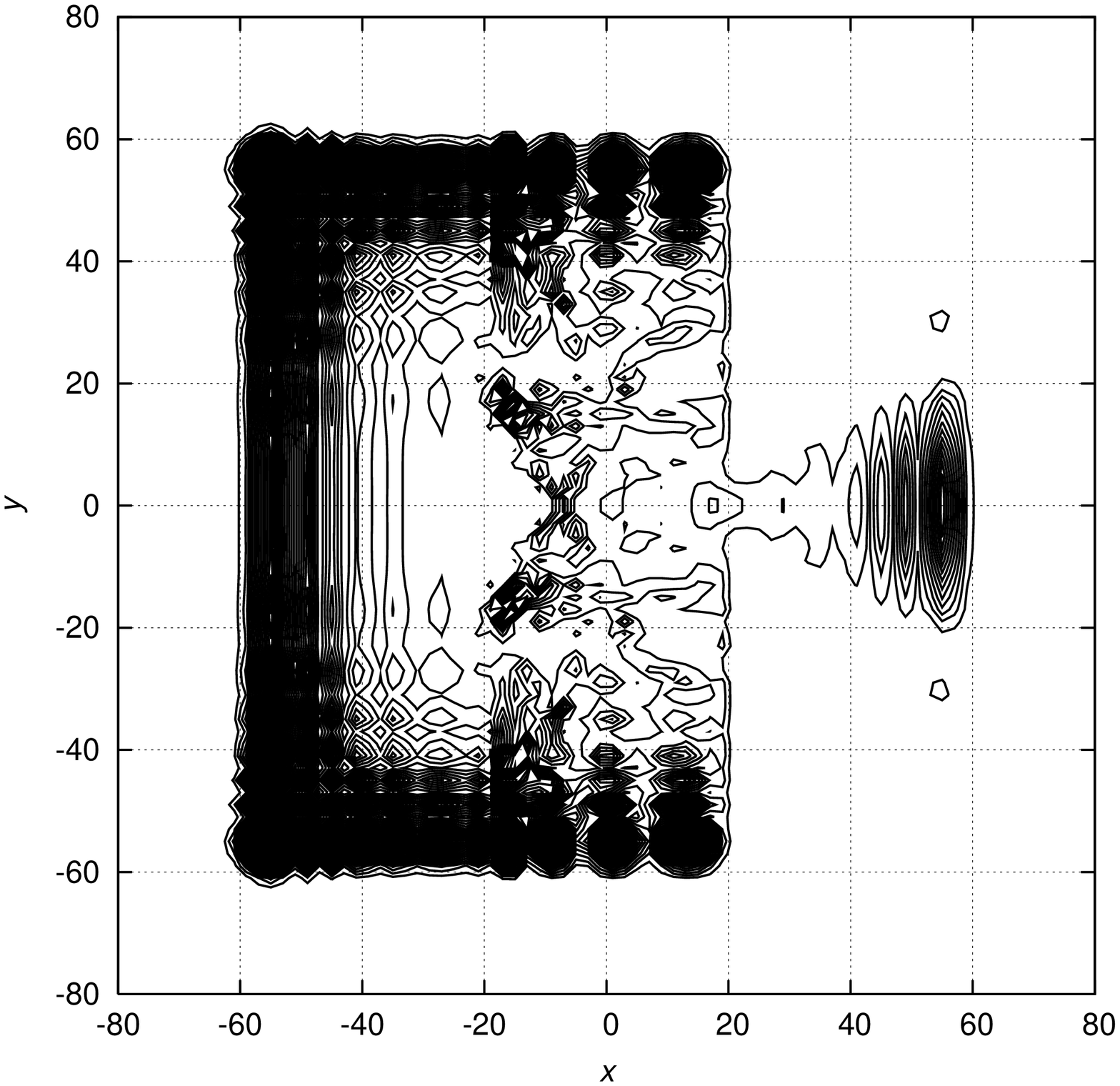}%
    \caption{{Probability distribution of the Hadamard walker
      after 80 iterations. The slit has a width of 5 units, and the
      barrier is placed at $x=20$.
      }} %
    \label{HadOneSlit}
\end{figure}

The simulation of a quantum walker passing through one slit
displays many interesting physical properties. Some of those are
similar to the ones that appear in the standard diffraction of
waves passing through one slit. In Fig.~\ref{HadOneSlit}, one can
see diffraction patterns after the Hadamard walker has passed
through one slit five units wide \footnote{The length unit is
defined as the distance between two neighboring sites in the
direction $x$ or $y$.}. Around $x=56$ there is a dominant
wavefront, which has a smooth central peak surrounded by two
secondary small peaks. The central peak is produced by
constructive interference and has a nature different from the
usual pattern of the Hadamard walker, which has the highest peaks
at the corners, not at the center of the side~\cite{Oliveira}. The
peaks at $x=-56$ are similar to those appearing for in the free
Hadamard walker, as it could be expected. The peaks between
$x=-20$ and $x=20$ are produced by the walkers reflected at the
barrier. As discussed below, besides these similarities in the
diffraction and interference patterns of quantum walkers and the
standard single-slit interference experiment, there are also some
noticeable differences.

\begin{figure}[h]
     \includegraphics[width=6cm,angle=270,scale=1]{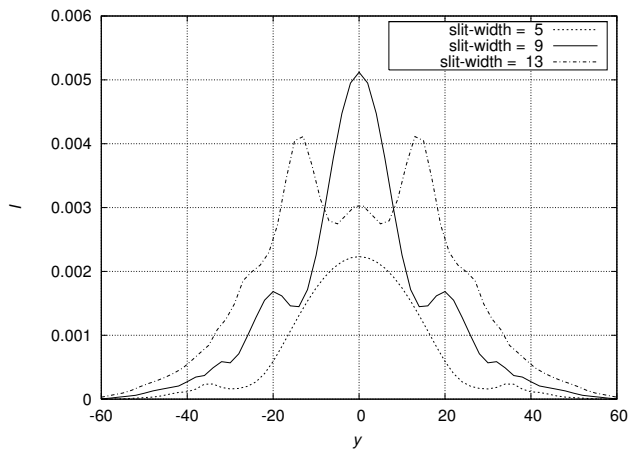}%
    \caption{{The probability distribution accumulated at the screen of the Hadamard
    walker. The number of iterations is 100 and the slit widths are 5, 9, and 13.
      The barrier is placed at $x=20$ and the screen is positioned at $x=60$. }}
    \label{HadScreenOneSlit}
\end{figure}

Fig.~\ref{HadScreenOneSlit} shows the probability distribution
accumulated at the screen ($I$) from $t=0$ until $t=100$ for the
Hadamard walker. We use the term accumulated in the following
sense: For each value of $y$ at the screen, we add up the values
of the probability from $t=0$ until $t=100$. At $t=100$, the
dominant contribution of the wave has already passed through the
screen. Note that in the previous figure, at $t=80$ the wavefront
has just arrived at the screen, which is located at $x=60$. In
Fig.~\ref{HadScreenOneSlit}, each curve corresponds to a slit
width (listed at the top). One can clearly see diffraction and
interference patterns. For each slit width, there is a central
peak produced by constructive interference and one local minimum
at each side produced by destructive interference. There are also
secondary maxima and minima. We found interesting to obtain these
patterns, as it is well known that the quantum walker Fourier
spectrum contains many different wavelengths~\cite{Nayak}. The
similarities with the standard single-slit diffraction pattern are
due to the dominance of particular wavelengths in the quantum
walker spectrum, and the differences to the multiplicity of those
wavelengths. The value of the intensity of the central peak
increases when the width increases until 9, but for larger widths,
the central peak decreases, gradually changing the pattern into
the usual Hadamard peaks at the extrema without intermediate
diffraction effects.

\section{The Double-Slit Case}\label{sec:2slits}

\begin{figure}[h]
    \includegraphics[width=5.5cm,angle=270,scale=1]{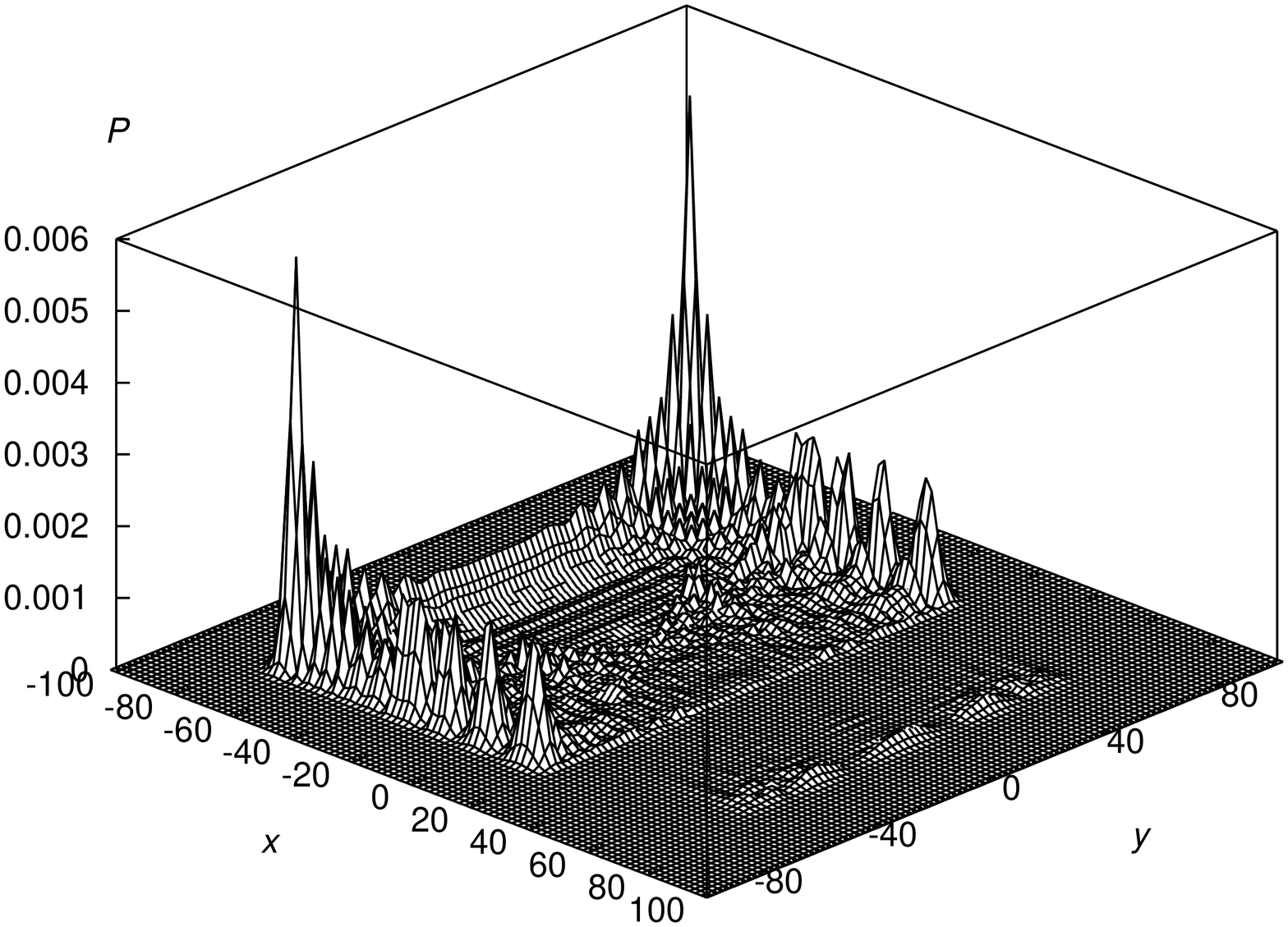}
    \includegraphics[width=5.5cm,angle=270,scale=1]{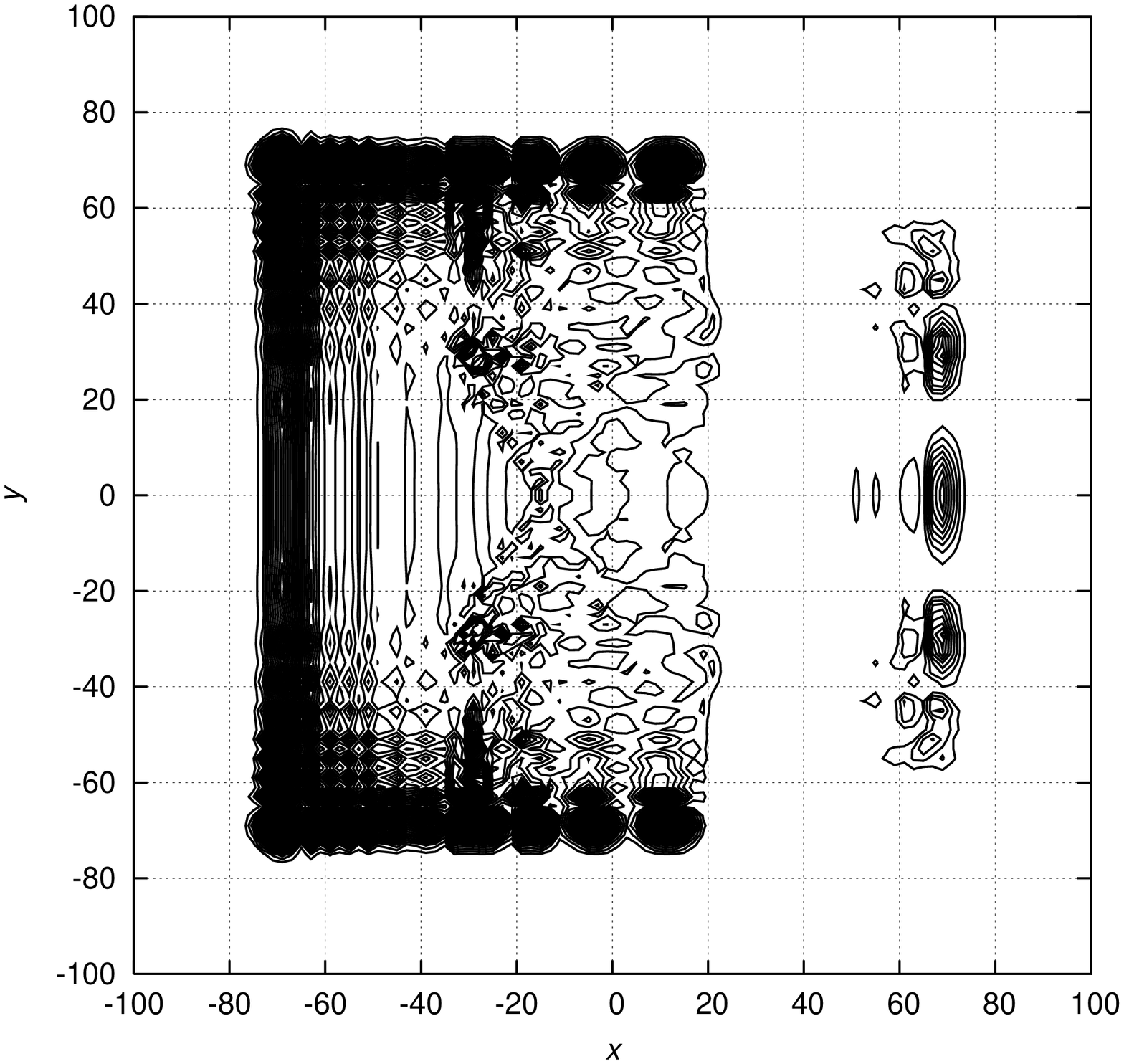}%
    \caption{{The probability distribution of the Hadamard walker
      after 100 iterations after having passed through a barrier positioned at $x=20$ with
      two slits of width 1 centered at $y=6$ and $y=-6$.
      The probability distribution of the diffracted wave
      ($x>20$) is amplified by a factor of 5.}} %
      \label{HadTwoSlits}
\end{figure}

In the Young's double-slit experiment, the intensity pattern at
the screen displays fringes produced by the interference of the
waves coming from each slit. The fringes are modulated by the
diffraction and interference due to the finite size of the slits.
If the slit width is small compared to the wavelength of the
light, the fringes are modulated by one large central peak. A
similar kind of pattern can be seen in Fig.~\ref{HadTwoSlits} for
the Hadamard walker. We have amplified the probability
distribution of the wave that has passed through the slits by a
factor of 5. The figure shows diffraction and interference
patterns. There are five peaks produced by constructive
interference interlaced by four valleys produced by destructive
interference. The wavefront is spread, as we have said before, due
to the small size of the slits compared to the order of the
smallest wavelengths present in the Fourier spectra of the quantum
walker. Note that the wavefront is flat and located around $x=70$.
In Young's experiment, the wavefront is a semi-circle. This
flatness is a characteristic of the Hadamard walker, produced by
quantum interference effects. Simulations with Grover and Fourier
coins yield a curved wavefront. Another difference from Young's
experiment is that the wavefront width is limited by the extrema
$y=100$ and $y=-100$. Outside this boundary, the wave function is
exactly zero.

\begin{figure}[h]
    \includegraphics[width=6cm,angle=270,scale=1]{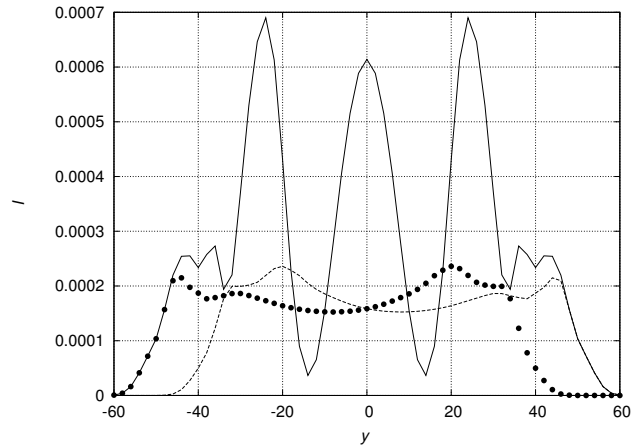}
    \caption{{The probability distribution accumulated at the screen
    of the Hadamard walker in the same settings of the previous figure.
    The screen is positioned at $x=60$. The dashed curve is the
    probability distribution when the lower slit is closed and upper
    slit is open. The dotted curve is the inverse case.}} %
      \label{HadScreenTwoSlits}
\end{figure}

Fig.~\ref{HadScreenTwoSlits} shows the distribution of probability
accumulated at the screen from $t=0$ until $t=100$ for the
Hadamard walker in the same settings of Fig.~\ref{HadTwoSlits}. It
is possible to see in details the sequence of alternating maxima
and minima inside a large envelope. The figure also depicts the
probability distributions accumulated at the screen when one slit
is closed and the other is open (dotted and dashed lines). Note
that if one adds up these curves, the result does not yield the
pattern of interference fringes. This result shows the wavelike
behavior of the walker. Note that the intensity at the first
minimum is close to zero, showing that the waves that come from
the slits have high coherence.

\begin{figure}[h]
    \includegraphics[width=6cm,angle=270,scale=1]{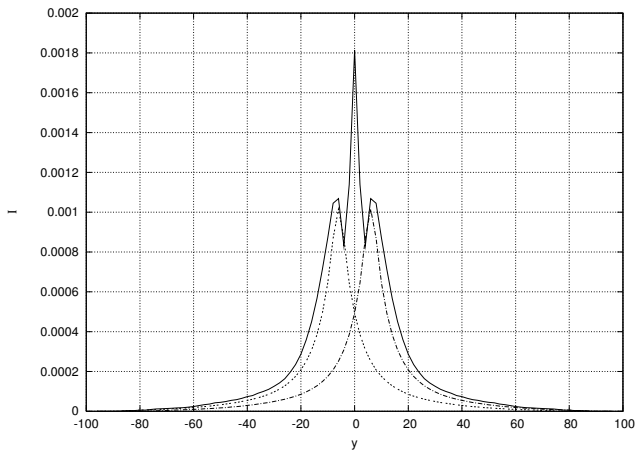}
    \caption{{The probability distribution accumulated at the screen
    of the Grover walker after 120 iterations after having
    passed through two slits at $x=30$ and screen at $x=70$.
    The positions of the slit centers are $y=6$ and $y=-6$.
    The slit width is 1.}} %
      \label{GroverScreenTwoSlits}
\end{figure}

It is interesting to analyze the physical behaviour for the Grover
and Fourier coins. Fig.~\ref{GroverScreenTwoSlits} shows the
probability distribution accumulated at the screen of the Grover
walker after having passed through two slits. The figure also
shows the distributions of probability when one slit is closed and
the other is open. Note that these isolated curves do not display
interference fringes. They consist of one peak. When both slits
are open, the interference pattern is characterized by a sharp
central peak surrounded by two smaller peaks, one at each side.
Note that the intensity at the minimum points is high. One can
understand this fact by analyzing the oscillating pattern (many
wavelengths) of the Grover wavefront at the $x$-axis, for example
in Ref.~\cite{Oliveira}. The oscillations are smoother along the
main diagonal, where one finds a single characteristic wavelength.
In fact, by putting the slits on a barrier perpendicular to the
main diagonal, one gets minimum points with low intensity similar
to the Hadamard walker. On the other hand, in all simulations with
the Grover walker we found three peaks, while for the Hadamard
walker we have obtained up to 9 peaks. Our simulations show that
the Fourier walker behaves in a way similar to the Grover walker.

\section{Conclusions}\label{sec:conc}

We have analyzed the diffraction and interference patterns of the probability distribution
of quantum walkers passing through one and two slits. We have used the evolution equation
of quantum walkers in two-dimensional lattices with broken links. The barrier with the slits
are generated by breaking permanently the appropriate links. The general setup is similar
to the one in the Young's double-slit interference experiment. We have performed simulations
using the Hadamard, Grover, and Fourier coins. The initial condition for the position is the
lattice center and for the coin is the one with maximum spreading rate.

In all simulations, diffraction and interference patterns were generated at the screen.
As expected, there are similarities and differences between the quantum walker diffraction and
interference patterns and the Young-experiment patterns. The similarities between these physical
phenomenon displays the wavelike behavior of quantum walkers. The differences can be explained
by the following facts. (1) It is known that the Fourier spectrum of the one-dimensional Hadamard
walker contains a plethora of wavelengths~\cite{Nayak}. The same is possibly true for Grover and
Fourier coins. In contrast, in Young's interference experiment, the photons have a single wavelength.
(2) The quantum walker evolution operator is different from the one used for free-moving fotons.

In future works, we are interested to introduce partial measurements to analyze the complementarity
relation between the appearance of interference fringes and the knowledge of which slit the walker
has passed. This analysis goes in the direction of Brukner and Zeiliger's work~\cite{Brukner} on the
finiteness of information in Young's experiment. Besides, the possibility of arbitrarily selecting
the shape of the region where the quantum walkers propagate, allows to verify the properties of these
systems in a variety of relevant configurations, e.g. chaotic billiards or moving walls.

\section*{Acknowledgments}
We thank F.L. Marquezino and G. Abal for useful discussions. This
work was funded by FAPERJ and CNPq.


\begin{thebibliography}{99}
\bibitem{Aharonov} Y. Aharonov, L. Davidovich, and N. Zagury,
\textit{Phys. Rev. A} \textbf{48}, 1687-1690 (1993).

\bibitem{Farhi} E. Farhi and S. Gutmann, {\em Phys. Rev. A} \textbf{58}, 915-928 (1998).

\bibitem{Shenvi} N. Shenvi, J. Kempe, and K. BirgittaWhaley,
\textit{Phys. Rev. A} \textbf{67}, 052307 (2003).

\bibitem{Ambainis} A. Ambainis, {\em Proc. 45th Symp. Found. Comp.
Sc.}, IEEE Computer Society Press, pp. 22-31, New York, 2004.

\bibitem{Nayak} A. Nayak and A. Vishwanath, arXiv:quant-ph/0010117 (2000).

\bibitem{Oliveira} A.C. Oliveira, R. Portugal, and R. Donangelo,
{\em Phys. Rev. A} \textbf{74}, 012312 (2006).

\bibitem{Romanelli} A. Romanelli, R. Siri, G. Abal,  A. Auyuanet,
and R. Donangelo, \textit{Physica A}, \textbf{347C},  137-152
(20056), also in arXiv:quant-ph/0403192.

\bibitem{Born} M. Born and E. Wolf, {\em Principles of Optics}, Pergamon
Press, Oxford, 1980.

\bibitem{Feynman} R.P. Feynman, R.B. Leighton, and M. Sands,
\textit{The Feynman Lectures on Physics}, Vol. 3, Addison-Wesley,
Boston, 1965.



\bibitem{Brukner} \v{C}. Brukner and A. Zeilinger, \textit{Phil. Trans. R. Soc.
Lond. A} \textbf{360}, 1061 (2002), e-print
arXiv:quant-ph/0201026.


\end{thebibliography}
\end{document}